\documentclass{ws-fnl}
\usepackage{hyperref}
\usepackage{cite}
\usepackage{amsmath}
\usepackage{graphicx}

\begin{document}

\markboth{Olha Shchur and Alexander Vidybida}{Firing statistics of spiking neuron with delayed feedback}

\catchline{}{}{}{}{}


\title{\bf First passage time distribution for spiking neuron with delayed excitatory
feedback}

\author{\footnotesize OLHA SHCHUR and ALEXANDER VIDYBIDA}

\address{Bogolyubov Institute for Theoretical Physics, Metrologichna str., 14-B,\\
Kyiv, 03680,
Ukraine\\
vidybida@bitp.kiev.ua}

\maketitle

\begin{history}
\received{(received date)}
\revised{(revised date)}
\end{history}

\begin{abstract}
A class of spiking neuronal models with threshold 2 is considered. It is defined by a set of conditions
typical for basic threshold-type models, such as the leaky integrate-and-fire (LIF)
or the binding neuron model and also for some artificial neurons.
A neuron is stimulated with a Poisson stream of excitatory impulses. Each output impulse is conveyed through the feedback line to the neuron input after 
finite delay $\Delta$. This impulse is identical to those delivered from the input stream.
 We have obtained a general relation allowing calculating exactly the 
probability density function (PDF) $p(t)$ 
for distribution of the first passage time of crossing the threshold, which is the
distribution 
of output interspike
 intervals (ISI) values for this neuron. 
 The calculation is based on known PDF $p^0(t)$
  for that same neuron without feedback, intensity of the input stream $\lambda$ and properties of the feedback line.
Also, we derive exact relation for calculating the moments of $p(t)$ based on
known moments of $p^0(t)$.

The obtained general expression for $p(t)$ is checked numerically using Monte Carlo simulation
for the case of LIF model.
The course of $p(t)$ has a $\delta$-function type peculiarity. This fact contributes to the discussion
about the possibility to model neuronal activity with Poisson process, supporting the ``no''
answer.

\noindent
{\bf Keywords.} spiking neuron; Poisson stochastic process; probability density function; delayed feedback; interspike interval statistics
\end{abstract}

\section{Introduction}

Activity of neurons either in the brain or at the periphery is highly irregular and is normally treated as random,\cite{Liley1956,Segundo1963,Gerstein1964,Smith1965,Segundo1966,Griffith1966a,Korolyuk1967,Drongelen1978,
Donnelly2004,Maimon2009,Averbeck2009}. 
For description/modeling of this activity simple stochastic processes are used, like
Poisson \cite[Fig.12]{Liley1956},\cite{Griffith1966a,Drongelen1978,Donnelly2004}, gated Poisson \cite{Smith1965}, Erlang \cite{Korolyuk1967}. 
It seems that a Poisson-type statistics is adequate at the periphery of the nervous system,
e.g. at the neuromuscular junctions, \cite{Liley1956}, the olfactory receptor neurons, 
\cite{Drongelen1978}, the chemoreceptor neurons, \cite{Donnelly2004}, the mechanoreceptor 
 neurons, \cite{Korolyuk1967}. For the primary visual cortex, already in 1966, it was observed that
 activity of only 1/7th portion of recorded neurons could be modeled by Poisson process, \cite{Griffith1966a}. 
In the more central areas, like the parietal lobe, neuronal activity exhibits a remarkable regularity
incompatible with Poisson-type statistics, \cite{Maimon2009,Averbeck2009}.

Deviation from Poisson statistics manifests itself in the deviation of the interspike intervals
probability distribution from the exponential form.
Many different reasons may bring on such a deviation.
The most evident reason is the requirement to have more than a single input impulse
for generating a single output spike. This makes the shortest ISIs less probable, see Fig. \ref{examples},{\em a}, Fig. \ref{trains}, {\em a,b}.
Other straightforward reasons might be a variable input, \cite{DOnofrio2016,Pirozzi2017},
or adaptation, \cite{Pirozzi2017}. One more reason is feedback, either positive or negative.
Such a feedback is often present if a neuron is embedded into a neuronal network
with non-zero transmission times, capable of reverberating dynamics. 
In this case, the feedback is normally mediated by other neurons and this may have a pronounced 
effect on the firing statistics of any of them. 

Our purpose is to check what might be that effect.

\begin{figure} [h]
\unitlength=0.9mm
\begin{center}
        \includegraphics[width=0.7\textwidth]{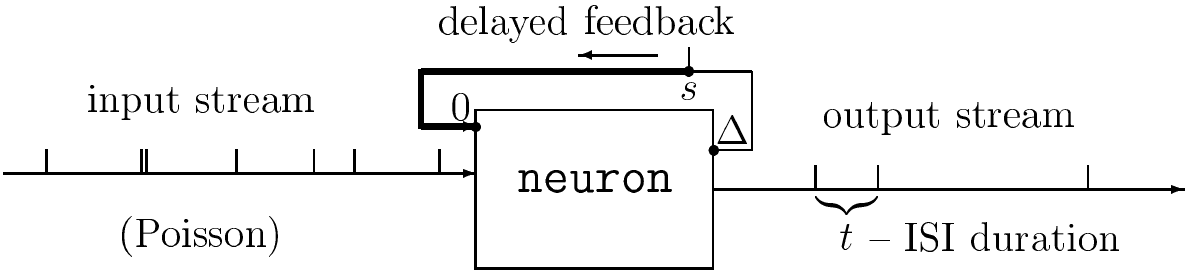}
\end{center}
\caption{\label{BNDwF} Neuron with delayed feedback. As {\large\tt neuron} in the figure,
we consider any neuronal model, which satisfies the set of conditions Cond0 - Cond5, above.}
\end{figure}

Rigorous mathematical treatment of stochastic behavior of neurons in a network with delayed
connections is a difficult task. In order to obtain an exact result, we take the simplest
possible case of feedback, namely, a neuron sends its output impulses onto its input, see Fig.
\ref{BNDwF}, below. This can be considered as the simplest, ``caricature network'' 
with delayed feedback, composed of a single neuron. It was our surprise that this kind of neuronal
organization has been observed experimentally, see Sec. \ref{Bjust}.

In this paper, we consider a neuron with {\em delayed excitatory} feedback. The purpose of this work
is to determine which effect the delayed feedback of this type might have on the 
probability density function (PDF) of output interspike intervals (ISI).
Mathematically, the problem consists of calculation of PDF $p(t)$  for 
the first passage times of crossing the threshold 
for a neuron
with the excitatory feedback, provided that the  PDF of the same type for the neuron without feedback, $p^0(t)$, and the properties of the feedback line are given.

It is worth noticing that in the  paper \cite{Vidybida2015a} the similar aim has been achieved but in the case of {\em instantaneous} excitatory feedback. As a result, 
the relation between three PDFs, namely for stimulus, for 
output stream without feedback and for output stream with instantaneous feedback, has been obtained.
This relation is used farther in this paper.

In our previous paper \cite{Vidybida2018},
we have considered the case of delayed fast {\em inhibitory} feedback.
We have obtained the relation between the PDF
for the neuron with delayed fast inhibitory feedback stimulated with Poisson stream
 and the PDF for the same neuron without 
feedback. In that work, we have come to the conclusion that the presence of delayed fast inhibitory feedback results in jump discontinuity of PDF at the point where ISI value equals  the delay in 
the feedback line. 
In \cite{Vidybida2018}, the problem has been stated for a whole class of threshold-type neuronal models that are determined through specifying a threshold level,
a height of the input impulses, and a time course of excitation decay. 
It has been shown there that for any
neuronal model (with and without feedback) that satisfies the imposed conditions there is invariant with respect to the way of excitation decay the initial segment of PDF. It is completely determined by the input stream and a ratio between the threshold level and the height of input impulses. In addition to this, it has appeared that the initial segment found for PDF for a neuron with feedback $p(t)$ is enough to express statistical moments of $p(t)$
through moments of $p^0(t)$, provided the time delay in the feedback line is shorter than the length
of the above-mentioned initial segment.

In this work,
we study the case of excitatory feedback with a non-zero delay ideologically similarly 
to our preceding work \cite{Vidybida2018} dealing with delayed fast inhibitory feedback. 
Biological justification of
this case of feedback is given in Sec. \ref{Bjust}. We consider a subclass of neuronal
models of the class considered in \cite{Vidybida2018}, which is 
composed of models having the firing 
threshold 2 (see Sec. \ref{class}, below). That is, the imposed in \cite{Vidybida2018} conditions on neuronal models, are valid here, see Cond0-Cond4, below.
 We assume that the neuron is stimulated with the input Poisson stream of excitatory impulses.
For this type of stimulation, we derive a general relation between the ISI values PDF $p(t)$
for a neuron with feedback based on known PDF $p^0(t)$ for the same neuron without feedback, see Eqs. (\ref{pif}), (\ref{a}),(\ref{g}) and (\ref{ptexc}).

Further, as it was mentioned previously, 
that for any neuronal model satisfying imposed conditions, there exists initial interval $]0;T_2[$ of ISI values at which $p^0(t)$ and, consequently, $p(t)$
does not depend on the neuronal model chosen. We calculate exactly
that model-independent initial segment of $p(t)$, see Sec. \ref{inv}.
As it can be clearly seen in Fig. \ref{examples} {\em b,c}, the $p(t)$ has peculiarity at the point equal to the delay time.

Moreover, we obtain in our approach the model-independent relations between the moments of PDF
of a neuron with and without feedback, see (\ref{moments}), (\ref{mean}), (\ref{w2}).  The first moment, the mean ISI value, was known before for the binding
neuron model only, see Eq. (\ref{mean}). The general expressions obtained here confirm what has been found previously for the binding neuron.

Finally, the exact expressions found
for $p(t)$ are verified by means of Monte Carlo simulation for 
the LIF neuron model, see Fig. \ref{examples} {\em b,c}.

\section{Methods}
\subsection{Class of neuronal models}\label{class}
We consider a class of neuronal models (without feedback) with threshold 2 stimulated with Poisson stream of input impulses of intensity $\lambda$. This class is a subset of a class of neuronal models considered in our preceding work \cite{Vidybida2018}, where any value for the threshold was considered.
The class considered here includes the leaky integrate-and-fire (LIF) and the binding
neuron  models,\footnote{Definition of the binding neuron model can be found in \cite{Vidybida2014}. See also\\ https://en.wikipedia.org/wiki/Binding\_neuron.} both with threshold 2.

The neuronal state at the moment $t$ is described by the depolarization voltage
$V(t)$, which considered to be biased by the resting potential value. Thus 
$V=0$ at the resting state and depolarization voltage is positive. 

The input impulse increases the depolarization
voltage by $h$:
\begin{equation}\label{h}
V(t)\quad\to\quad V(t)+h,
\end{equation}
where $h>0$.

The input impulse decay is determined by a decreasing function $y(u)$,
which is different for different neuronal models. It means that
if an impulse is received at the moment $t$, then for any $u>0$
\begin{equation}\label{decay}
V(t+u)=v(t+u)+hy(u),
\end{equation}
where $v(t+u)$ denotes depolarization voltage without that impulse involved.

For example, for the LIF model one has
\begin{equation}\label{fLIF}
y(u)=e^{-\frac{u}{\tau}},
\end{equation}
where $\tau$ is the relaxation time.

The neuron is characterized by a firing threshold value $V_0$: as soon as $V(t)>V_0$,
the neuron generates a spike and $V(t)$ becomes zero.

Instead of specifying any concrete neuronal model (through specifying $V_0$, $h$ and $y(u)$), we consider a class of neuronal models, which (without feedback) satisfy the following conditions:
\begin{itemize}
\item Cond0: Neuron is deterministic: Identical stimuli elicit identical 
spike trains from the same neuron.
\item Cond1: Neuron is stimulated with input Poisson stream of excitatory impulses.
PDF of intervals between those impulses is 
\begin{equation}\label{pin}
p^{in}(t)=\lambda e^{-\lambda t},
\end{equation}
where t means an input ISI duration.
\item Cond2: Just after firing, neuron appears in its resting state.
\item Cond3: The function $y(u)$, which governs decay of excitation, 
see Eq. (\ref{decay}), is continuous and satisfies the following conditions:
\begin{equation}\label{decrease}
y(0)=1,\qquad
0<u_1<u_2\quad\Rightarrow\quad y(u_1)\geq y(u_2).
\end{equation}
\item Cond4: The PDF for output ISIs, $p^0(t)$, where $t$ means an output ISI duration, exists 
together with all its moments.
\item Cond5: Neuron has a threshold 2. It means that in order to be triggered neuron should
obtain at least 2 impulses, or, in other words, the following relation between the threshold $V_0$ and the input impulse height $h$ is fulfilled:
\begin{equation}
1< \dfrac{V_0}{h}<2.
\end{equation}
\end{itemize}
Notice, that Cond0-4, above, are the same as in \cite{Vidybida2018}.

As it has been shown in \cite{Vidybida2018}, for any neuronal model that satisfies Cond0-5 there is the initial segment of PDF $p^0(t)$ of length $T_2$ where the PDF 
 $p^0(t)$ on that segment looks as follows\footnote{In the (\ref{p0}), 
 we have taken into account that the threshold equals two.}
\begin{equation}\label{p0}
p^{0}(t) dt=\lambda t e^{-\lambda t} \lambda dt,\:t<T_2.
\end{equation} 
The concrete value of $T_2$ is determined by a way of excitation decay (the explicit form of $y(u)$, 
see \cite[Sec. 3.3]{Vidybida2018}).

\subsection{Type of feedback}
To the neuronal model which satisfies the Cond0-Cond5 mentioned above we add the delayed excitatory feedback line and expect that it has the following properties:
\begin{itemize}
\item Prop1: The time delay in the line is $\Delta>0$.
\item Prop2: The line is able to convey no more than one impulse.
\item Prop3: The impulse conveyed to the neuronal input is excitatory impulse identical
to those from the input stream.
\item Prop4: The delay in the feedback line $\Delta$ satisfies the following condition:\begin{equation}\label{condDT}
\Delta < T_2.
\end{equation}
\end{itemize}
Notice, that Prop1-2, above, are the same as in \cite{Vidybida2018}.

\subsection{Biological justification}\label{Bjust}
In the natural conditions, neurons are embedded into reverberating networks. Since
output neuronal activity can be fed back transformed and transmitted
by other neurons,
most neurons can “feel” their own activity produced some time earlier. This justifies consideration of a feedback in neural systems.

At the same time, it is known that excitatory neurons can form synapses (autapses) on its
own body or dendritic tree \cite{Bekkers1998,Aroniadou-Anderjaska1999}. This substantiates consideration of a single neuron with excitatory feedback not only as the simplest reverberating “network” possible but also as an
independent biologically relevant case. The delay $\Delta$ comprises the time required by the output spike to pass the distance from axonal hillock, where it is generated, to the autapse and
the synaptic delay.
\section{Results}
\subsection{Pdf: general relation}
The strategy for finding the PDF for a neuron with delayed excitatory feedback is the same as in our previous work \cite{Vidybida2018} on delayed fast inhibitory feedback.

Let us introduce a time-to-live $s$ --- the time at the beginning of ISI needed for an
impulse in the feedback line to reach a neuron. $f(s)$ is a distribution of such 
times-to-live in a stationary regime. If $p(t|s)$ is the conditional PDF that allows calculating the probability to obtain an ISI duration $t$ given
that the time-to-live at the beginning of the ISI is $s$, then in case of stationary regime
the wanted PDF for a neuron with feedback can be found in the following way:
\begin{equation}\label{pt}
p(t)=\int\limits_0^\Delta ds\: p(t|s) f(s).
\end{equation}

The conditional PDF $p(t|s)$ depends on the type of feedback.
Now we need to figure out what $p(t|s)$ looks like  in case of excitatory feedback.

\subsubsection{Conditional probability $p(t|s)$}
Below we will scrutinize different cases for the sake of obtaining the exact expression
for $p(t|s)$. Firstly, the probability to get an ISI $t$ shorter than time-to-live
$s$ does not depend on the feedback line presence and is the same as in the case of the feedback absence. Thus
\begin{equation}
t<s \Rightarrow p(t|s)\:dt=p^0(t)\:dt.
\end{equation}

Secondly, there is nonzero probability to obtain ISI with exact duration $t=s$. In order to
get such an ISI duration, a neuron must receive one input impulse before receiving an impulse from the
feedback line. The condition (\ref{condDT}) ensures that in this case at the moment
$t\in]s-\epsilon;s+\epsilon[$ there will be enough excitation in the neuron to surpass the
threshold. The probability of this event, i.e. to obtain one impulse (and no more) from input Poisson stream
on time interval $]0;t[$, is $\lambda t e^{-\lambda t}.$
Therefore for the conditional probability $p(t|s)$ in this case:
\begin{equation}
t\in]s-\epsilon;s+\epsilon[ \Rightarrow p(t|s)ds=\lambda t e^{-\lambda t} \delta(t-s)ds.
\end{equation}

Thirdly, if we obtain an ISI duration $t>s$, it means that before 
receiving an impulse from the feedback line, due to 
(\ref{condDT}), there has not been any excitation on the neuron, i.e. the neuron
has not received an impulse from the input line. The probability of this event is simply 
$e^{-\lambda s} $. Right away after the receiving that impulse, the neuron 
is in the same state as at 
the beginning of ISI if there was an instantaneous feedback instead of delayed. 
Let us denote PDF for the same neuron but with instantaneous feedback (i.e. when $\Delta=0$) as $p^{o\_if}(t)$.
Thus, in this case, we have the following:
\begin{equation}
t>s \Rightarrow p(t|s)\:dt=e^{-\lambda s} p^{o\_if}(t-s)\:dt.
\end{equation}

In \cite{Vidybida2015a} the relation between PDF for a neuron without feedback $p^0(t)$
and PDF for the same neuron with instantaneous feedback $p^{o\_if}(t)$, stimulated  with Poisson 
stream, has been derived. 
It looks as follows:
\begin{equation}\label{pif}
p^{o\_if}(t)=p^0(t)+\dfrac{1}{\lambda}\dfrac{d}{dt}p^0(t).
\end{equation}

To summarize, the conditional probability $p(t|s)$ can be written in the 
following form:
\begin{equation}\label{pts}
p(t|s)=\chi(s-t)p^0(t)+\lambda t e^{-\lambda t}\delta(t-s)+
e^{-\lambda s}p^{o\_if}(t-s),
\end{equation}
where $\chi(t)$ is the Heaviside step function.

\subsubsection{Distribution of times-to-live $f(s)$}
As regards the distribution of times-to-live $f(s)$, in \cite{Vidybida2018}, it was proved
that $f(s)$ has the following form:
\begin{equation}\label{fs}
f\left(s\right)=g\left(s\right)+a\delta\left(s-\Delta\right),
\end{equation}
where 
\begin{equation}\label{a}
a=\dfrac{4e^{2\lambda\Delta}}{1+e^{2\lambda\Delta}(2\lambda\Delta+3)},
\end{equation}
and
\begin{equation}\label{g}
g(s)=\dfrac{a\lambda}{2}\left(1- e^{-2\lambda(\Delta-s)}\right).
\end{equation}

\subsubsection{General relation for PDF $p(t)$}
After substituting (\ref{pts}) and (\ref{fs}) into ({\ref{pt}), one can obtain the
following general formulae for different domains of ISI duration:
\begin{equation}\label{ptexc}
p(t)=\begin{cases}
\int\limits_0^t ds\: e^{-\lambda s} p^{o\_if}(t-s) g(s)+\\
+p^0(t)(\int\limits_t^\Delta ds\: g(s)
+a)+g(t)\lambda t e^{-\lambda t},&t< \Delta;	\\
a\lambda \Delta e^{-\lambda \Delta}\delta(t-\Delta),
&t\in]\Delta-\epsilon;\Delta+\epsilon[;\\
a e^{-\lambda \Delta} p^{o\_if}(t-\Delta)+\\
+\int\limits_0^\Delta ds\: e^{-\lambda s} p^{o\_if}(t-s) g(s),&t>\Delta.
\end{cases}
\end{equation}

As it can be seen from right above, the PDF $p(t)$ has a Dirac $\delta$-function type peculiarity
at the point corresponding the ISI duration that equals the delay in the feedback line $\Delta$.

To sum up, we have the following algorithm for finding the PDF $p(t)$ for a neuron with feedback, if 
the PDF $p^0(t)$ for the same neuron without feedback is known:
\begin{enumerate}
\item find $p^{o\_if}(t)$ by using (\ref{pif});
calculate  $p(t|s)$;
\item substitute $p^{o\_if}(t)$ and components of the distribution function $f(s)$ (\ref{a}) and  (\ref{g}) into (\ref{ptexc}) and take integrals.
\end{enumerate}

\subsection{Model-invariant initial segment of $p(t)$}\label{inv}
We can find the initial segment $]0;T_2[$ of $p(t)$ by substituting (\ref{p0}), 
(\ref{pif}), (\ref{a}), (\ref{g}) into (\ref{ptexc}): 
\begin{equation}\label{initpt}
p(t)=\begin{cases}
\dfrac{\lambda e^{-\lambda t}}{3+2\Delta\lambda+e^{-2\lambda\Delta}}(e^{-2\lambda\Delta}
\left(1-e^{2\lambda t}(1+\lambda t)\right)+\\
\lambda t(7+2\Delta\lambda)-2\lambda^2 t^2),
&\text{$t<\Delta$;}\\
\dfrac{4\lambda \Delta e^{-\lambda\Delta}}{3+2\Delta\lambda+e^{-2\lambda\Delta}}
\delta(t-\Delta),&\text{$t\in]\Delta-\epsilon;\Delta+\epsilon[$;}\\
\lambda e^{-\lambda t},&\text{$t\in]\Delta;T_2[$.}
\end{cases}
\end{equation}

This is in agreement with \cite[Eq. (18-20)]{Vidybida2008a} and \cite[Eq. (12-14)]{Vidybida2009} where PDF has been obtained for the binding 
neuron model only. It is necessary to emphasize that in the present work the exact expression for $p(t)$ (\ref{initpt}) on the initial segment $]0;T_2[$ of ISI values
  is
the same for the whole class of neuronal models with threshold 2 stimulated with Poisson input stream
(invariant with respect to a way of excitation decay).

The validity of the obtained formulas  (\ref{initpt}) has been verified using
numerical simulation for the LIF model by means of the Monte Carlo method, see Fig. \ref{examples} {\em c}.

\begin{figure}

    \includegraphics[width=1.0\textwidth]{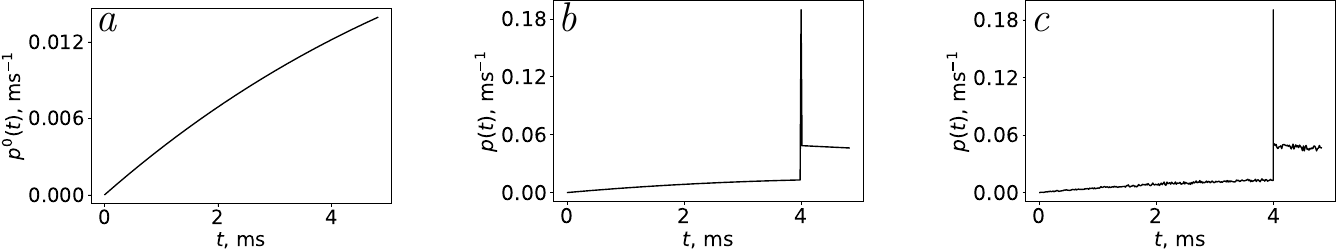}
  \caption{\label{examples}  Example of ISI PDF for the LIF neuron with threshold 2 without feedback ({\em a}, used Eq. (\ref{p0})) and with delayed excitatory feedback ({\em b}, used Eq. \ref{initpt})).
  For both panels:
$\tau=20$ ms, $V_0=20$ mV, $h=11.2$ mV, $\lambda=62.5$ s$^{-1}$. $\Delta=4$ ms in ({\em b}).
{\em c} --- Monte Carlo simulation for $p(t)$ (used 1\,000\,000 output ISIs).
 The total probability under the curves is 0.03726 ({\em a}) and 0.42205 ({\em b}).}\end{figure}

\subsection{Moments of PDF}\label{mome}
Let us denote the $n$th moment of PDF for a neuron with delayed feedback $p(t)$ and
without feedback $p^0(t)$ as $W_n$ and $W^0_n$ respectively. Using the definitions of
the moments $W^0_n$ and (\ref{ptexc}) within different time domains, one can obtain the following expression for the 
$n$th moment of PDF of ISI duration for a neuron stimulated with Poisson stream:
\begin{equation}\label{moments}
\begin{split}
W_n=
a\Delta^n \lambda \Delta e^{-\lambda \Delta} +
\int\limits_0^\Delta dt\:t^n \left(g(t)\lambda \Delta e^{-\lambda \Delta} + a p^0(t)+p^0(t)\int\limits_t^\Delta ds\:g(s)
\right)
\\+ \sum_{k=0}^n \binom{n}{k} (W_k^0-\frac{1}{\lambda}kW_{k-1}^0 )
\left(a e^{-\lambda \Delta} \Delta^{n-k}
+
\int\limits_0^\Delta ds\: g(s) e^{-\lambda s}s^{n-k}\right)
\end{split}
\end{equation}

$W^0_0$ is a normalization coefficient and equals $1$. Notice that the explicit expression of $p^0(t)$ in formula
right above is used only for values on interval $]0;\Delta[$ where it is given by (\ref{p0}).

The exact expression for the first moment of $p(t)$ looks as follows:
\begin{equation}\label{mean}
W_1=
\dfrac{2\left(-1+W_1^0\lambda+e^{2\Delta\lambda}(-1+W_1^0\lambda +2\Delta\lambda)\right)}
{\lambda(1+e^{2\lambda\Delta}(2\lambda\Delta+3))}.
\end{equation} 

The expression right above is in agreement with that found for binding neuron model in
 \cite{Vidybida2008a} and \cite{Vidybida2009}.
 
In the case of instantaneous feedback, i.e. when $\Delta=0$, for the mean of $p(t)$
we have:
\begin{equation}\label{w1if}
W_1^{0\_if}=W_1^0-\frac{1}{\lambda}.
\end{equation}

As regards the second moment, $W_2$, the exact expression is:
\begin{equation}\begin{split}\label{w2}
W_2=
\dfrac{2}{\lambda^2(1+e^{2\lambda\Delta}(2\lambda\Delta+3))}(4+\lambda-4W_1^0\lambda+W_2^0\lambda ^2\\+e^{2 \Delta  \lambda } \lambda  
(-1+W_2^0 \lambda +2 \Delta
\lambda )+2 e^{\Delta  \lambda } (-2-\Delta ^2 \lambda ^3+\Delta ^3 \lambda ^3))
.\end{split}
\end{equation}

In case of the instantaneous feedback the second moment is:
\begin{equation}\label{w2if}
W_2^{0\_if}=W_2^0-\frac{2}{\lambda}W_1^0.
\end{equation}

(\ref{w1if}) and (\ref{w2if}) are in concordance with \cite[Eq. (7),(8)]{Vidybida2015a}
 where relations between the first two moments
for a neuron with instantaneous feedback and for the same neuron without feedback
have been obtained.

\begin{figure}
    \includegraphics[width=1.0\textwidth]{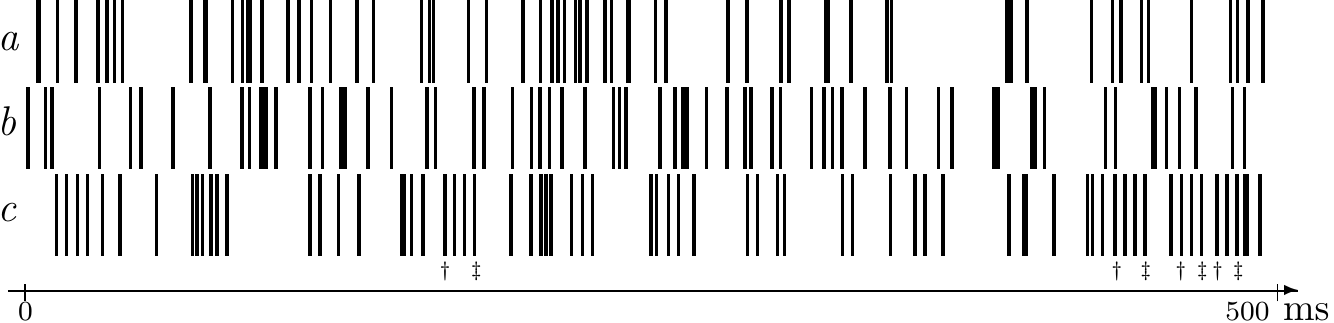}
  \caption{\label{trains} Examples of spike trains for the input Poisson stream
  with intensity $\lambda=130$ s$^{-1}$, {\em a}, the LIF neuron with threshold 2 without feedback ($\lambda=300$ s$^{-1}$), {\em b}, and with delayed excitatory feedback ($\lambda=220$ s$^{-1}$), {\em c}.
The $\lambda$ values are chosen in such a way that the total number of spikes at the 0 --- 500 ms
interval is the same. It may be observed that in {\em b} the spikes are distributed more evenly
than in {\em a}.
In {\em c}, notice the series of ISIs equal $\Delta$ in periods between $\dagger$ and
$\ddagger$.
  }
\end{figure}

\section{Conclusions and Discussion}

In the current paper, we study the effect of the presence of the delayed excitatory feedback line on the neuronal activity.
For a class of neural models with threshold 2 stimulated with Poisson process and satisfying the imposed conditions (see Sec. \ref{class}), we have derived the general relation
between the PDF $p(t)$ of ISI values for a neuron with excitatory feedback
and the corresponding PDF $p^0(t)$ for the same neuron, but without feedback, see Eqs. (\ref{pif}), (\ref{a}), (\ref{g}) and (\ref{ptexc}). Furthermore,
the initial model-invariant segment of PDF $p(t)$ has been found in the explicit form, see
(\ref{initpt}). Also,
we express the moments of PDF $p(t)$ for a neuron with feedback through the corresponding 
moments of PDF $p^0(t)$ for the neuron without feedback. The obtained relations between moments
must be met for any neuronal model with threshold 2 stimulated with Poisson stream (i. e. that satisfies the imposed conditions Cond0-5, above). 
Notice that we consider here a deterministic threshold. In reality, the exact value of the firing
threshold may fluctuate for different reasons. Results for fluctuating threshold obtained in
diffusion approximation can be found in \cite{Pirozzi3}.
All obtained results are mathematically rigorous.
Other rigorous results as regards statistics of neuronal activity can be found in the review \cite{Sacerdote2013}.

As it can be clearly seen in Fig. \ref{examples} {\em b,c}, the course of the PDF found has a  peculiarity --- a Dirac $\delta$-function 
for ISI $t=\Delta$, which makes it impossible to describe the output neuronal stream by a 
Poisson-like or other simple distribution.
If the output of a neuron is fed to another neuron, the spiking pattern of the output 
determines the reaction of the post-synaptic neuron, see e.g. \cite{Segundo1963,Segundo1966}.
The $\delta$-function presence in the $p(t)$ may rearrange the spiking pattern by adding
some regularity, as illustrated in Fig. \ref{trains} {\em c}.

\section*{Acknowledgments}
This research was supported by theme grant of the department of physics and
astronomy of NAS of Ukraine: "Dynamics of formation of spatially
non-uniform structures in many-body systems", PK 0118U003535.

\end{document}